\documentclass[12pt]{article}
\pdfoutput=1

\usepackage{amsmath,amssymb,amscd}
\usepackage{listings}
\usepackage{caption}
\usepackage{dsfont}
\usepackage{slashed}
\usepackage{color}
\usepackage{ulem}

\usepackage[pdftex]{graphicx}
\usepackage{epstopdf}
\usepackage{subfigure}
\usepackage{epsfig}
\usepackage{listings}
\usepackage{caption}
\usepackage{cite}

\usepackage{multirow}

\newcommand{\beq}{\begin{equation}}
\newcommand{\eeq}{\end{equation}}
\newcommand{\nbea}{\begin{align*}}
\newcommand{\neea}{\end{align*}}
\newcommand{\nbeq}{\begin{equation*}}
\newcommand{\neeq}{\end{equation*}}

 \usepackage{multirow}
\usepackage{array}
\newcolumntype{M}[1]{>{\centering\arraybackslash}m{#1}}
\newcolumntype{N}{@{}m{0pt}@{}}

\begin{document}


\pagestyle{empty}

\baselineskip=21pt
\rightline{{\fontsize{0.40cm}{5.5cm}\selectfont{KCL-PH-TH/2016-08, LCTS/2016-06, CERN-TH/2016-044}}}
\vskip 0.2in

\begin{center}

{\large {\bf Photon Mass Limits from Fast Radio Bursts}}

\vskip 0.2in


{\bf Luca Bonetti}\textsuperscript{a,b},~
{\bf John Ellis}\textsuperscript{c,d},~
{\bf Nikolaos E. Mavromatos}\textsuperscript{c,d},~\\
{\bf Alexander S. Sakharov}\textsuperscript{e,f,g},~
{\bf Edward K. Sarkisyan-Grinbaum}\textsuperscript{g,h},~\\
{\bf Alessandro D.A.M. Spallicci}\textsuperscript{a,b}

\vskip 0.2in

{\small {\it

\textsuperscript{a}{\mbox Observatoire des Sciences de l'Univers en r\'egion Centre, UMS 3116, Universit\'e d'Orl\'eans}\\
\mbox{1A rue de la F\'erollerie, 45071 Orl\'eans, France}\\  
\vspace{0.25cm}
\textsuperscript{b}Laboratoire de Physique et Chimie de l'Environnement et de l'Espace, UMR 7328\\
Centre Nationale de la Recherche Scientifique\\
\mbox{LPC2E, Campus CNRS, 3A Avenue de la Recherche Scientifique, 45071 Orl\'eans, France}\\
\vspace{0.25cm}
\textsuperscript{c}Theoretical Particle Physics and Cosmology Group, Physics Department, \\
King's College London, Strand, London WC2R 2LS, United Kingdom\\
\vspace{0.25cm}
\textsuperscript{d}Theoretical Physics Department, CERN, CH-1211 Gen\`eve 23, Switzerland\\
\vspace{0.25cm}
\textsuperscript{e}Department of Physics, New York University\\
4 Washington Place, New York, NY 10003, United States of America\\
\vspace{0.25cm}
\textsuperscript{f}Physics Department, Manhattan College\\
{\mbox 4513 Manhattan College Parkway, Riverdale, NY 10471, United States of America}\\
\vspace{0.25cm}
\textsuperscript{g}Experimental Physics Department, CERN, CH-1211 Gen\`eve 23, Switzerland \\
\vspace{0.25cm}
\textsuperscript{h}Department of Physics, The University of Texas at Arlington\\
{\mbox 502 Yates Street, Box 19059, Arlington, TX 76019, United States of America} \\}}

\vskip 0.2in

{\bf Abstract}

\end{center}

\baselineskip=18pt \noindent


{\small
The frequency-dependent time delays in fast radio bursts (FRBs) can be used to constrain the photon mass,
if the FRB redshifts are known, but the similarity between the frequency dependences of
dispersion due to plasma effects and a photon mass complicates the derivation of a limit on $m_\gamma$.
The dispersion measure (DM) of FRB 150418 is known to $\sim 0.1$\%, and there is a claim to have
measured its redshift with an accuracy of $\sim 2$\%, but the strength of the constraint on $m_\gamma$ is limited by
uncertainties in the modelling of the host galaxy and the Milky Way, as well as possible
inhomogeneities in the intergalactic medium (IGM). 
Allowing for these uncertainties,
the recent data on FRB 150418 indicate that
$m_\gamma \lesssim 1.8 \times 10^{-14}$~eV c$^{-2}$ ($3.2 \times 10^{-50}$~kg), if FRB 150418 indeed has a redshift
$z = 0.492$ as initially reported.
In the future, the different redshift dependences of the plasma and photon mass contributions to DM
can be used to improve the sensitivity to $m_\gamma$ if more FRB redshifts are
measured. For a fixed fractional uncertainty in the extra-galactic contribution to the
DM of an FRB, one with a lower redshift would provide greater sensitivity to $m_\gamma$.\\
\begin{center}
{\it We dedicate this paper to the memory of Lev Okun, an expert on photon mass}
\end{center}
}




\newpage
\pagestyle{plain}

When setting an upper limit on the photon mass,
the Particle Data Group (PDG)~\cite{PDG} cites the outcome of modelling the solar system magnetic field:
first at 1 AU, $m_\gamma < 5.6 \times 10^{-17}$~eV c$^{-2}$ ($=10^{-52}$~kg) \cite{ry97,ry07}, and later at $40$ AU,
$m_\gamma < 8.4 \times 10^{-19}$~eV c$^{-2}$ ($= 1.5 \times 10^{-54}$~kg) \cite{ry97}.
However, the laboratory upper limit is four orders of magnitude larger~\cite{wifahi71}; 
for reviews see \cite{OtherReviews,goni10}. In \cite{goni10}, the authors state the concern 
that ``{\it Quoted photon-mass limits have at times been overly optimistic in the strengths of their characterizations. 
This is perhaps due to the temptation to assert too strongly something one `knows' to be true"}.  
This concern was mainly addressed to the galactic magnetic field model limits~%
\cite{mgf}, but it should be borne in mind also when assessing the solar system limits. 

Indeed, the estimates on the deviations from Amp\`ere's law in the solar wind \cite{ry97,ry07} are not based simply on
{\it in situ} measurements. For example: (i) the magnetic field is assumed to be exactly, always and everywhere a 
Parker spiral; (ii) the accuracy of particle data measurements  from, {\it e.g.}, Pioneer or Voyager, has not been discussed; 
(iii) there is no error analysis, nor data presentation, instead; (iv) there is extensive use of a {\it reductio ad absurdum} 
approach based on earlier results of other authors, which are often devoted to other issues than establishing a 
basis for an extremely 
difficult measurement of a mass that is many orders of magnitude lower than that of an electron or a neutrino. 

In order to check these estimates of the solar wind at 1 AU, a more experimental approach has been pursued via a 
thorough analysis of Cluster data~\cite{respva2016}, leading to a mass upper limit lying between 
$1.4 \times 10^{-49}$ and $3.4 \times 10^{-51}$ kg, 
according to the estimated potential. The difference between the results of this conservative approach and previous estimates,
as well as the need for astrophysical modelling, motivate the
development of additional methods for constraining the photon mass.

The time structures of electromagnetic emissions from astrophysical sources at cosmological distances
have been used to constrain other aspects of photon/electromagnetic wave propagation,
such a possible Lorentz-violating energy/frequency dependence of the 
velocity of light {\it in vacuo}~\cite{aemns, stretch1, MAGIC, HESS, GRB2}, and the
possibility of dispersion in photon velocities of fixed energy/frequency, as suggested by some models of
quantum gravity and space-time foam~\cite{aemn, delays}. Similarly, the gravitational waves recently observed by Advanced LIGO
from the source GW150914 have been used to constrain aspects of graviton/gravitational wave propagation,
including an upper limit on the graviton mass: { $m_g < 1.2 \times 10^{-22}$~eV c$^{-2}$ 
($= 2.1 \times 10^{-58}$~kg)}~\cite{LIGO, LIGO2} and
limits on Lorentz violation~\cite{EMNGW, Kost}, and the possible observation by Fermi of an
associated $\gamma$-ray pulse~\cite{Fermi} suggests that light and gravitational waves have
the same velocities to within $10^{-17}$~\cite{EMNGW, others}. 

The time structures of electromagnetic emissions from astrophysical sources 
at cosmological distances can also be used to derive an upper limit on the photon mass, $m_\gamma$.
Since the effect of the photon mass on the velocity of light is enhanced at low frequency $\nu$ (energy $E$):
{$\Delta v \propto - m_\gamma^2c^4/h^2\nu^2$ ($- m_\gamma^2c^4/E^2$)}, measurements
of time structures at low frequency or energy are particularly sensitive
to $m_\gamma$. For this reason, measurements of short time structures in radio emissions
from sources at cosmological distances are especially powerful for constraining $m_\gamma$.
This is to be contrasted with probes of Lorentz violation, for instance, where measurements of
{ high-energy photons} such as $\gamma$ rays are at a premium. This is why probes of the photon
mass using {gamma-ray} bursters (GRBs)~\cite{GRBmg} and active galactic nuclei (AGNs) have not been
competitive in constraining $m_\gamma$. As we mention later, a stronger limit
can be obtained by using the apparent coincidence of a radio afterglow with a GRB, but this
is also not competitive with the sensitivity offered by fast radio bursts (FRBs).

FRBs are potentially very interesting because their radio signals have {well-measured}
time delays that exhibit the $1/\nu^2$ dependence expected for both the free electron density along the line of sight
and mass
effects on photon propagation. Until recently, the drawback was that no FRB had had
its redshift measured, though there was considerable evidence that they occurred at
cosmological distances. This has now changed with FRB 150418~\cite{FRB}, which has been reported
to have occurred in a galaxy with a well-measured redshift $z = 0.492 \pm 0.008$.
The identification of its host galaxy has been questioned, and the alternative possibility of a coincidence
with an AGN flare has been raised~\cite{AGN}, though the likelihood of this is currently an open question~\cite{debate}.
In the following we assume the host galaxy identification made in~\cite{FRB}, and also discuss more generally
how non-galactic FRBs could be used to constrain photon propagation.

The frequency-dependent time lag of FRB 150418 between the arrivals of pulses with $\nu_1=1.2$~GHz and
$\nu_2=1.5$~GHz is $\Delta t_{12}^{\rm FRB}\approx 0.8$~s,
and  was used in~\cite{FRB} to extract very accurately the dispersion measure (DM), which is given in the absence
of a photon mass by the integrated column density of free electrons along the propagation path of a radio 
signal, $\int n_edl$. The
delay of an electromagnetic wave with frequency $\nu$ propagating through a plasma with an 
electron density $n_e$, relative
to a signal in a vacuum, makes the following frequency-dependent contribution to the time delay~\cite{IGM,IGM2}
\begin{equation}
\Delta t_{\rm DM} \; = \; \int\frac{dl}{c}\frac{\nu_p^2}{2\nu^2} \; =\; 
415\left( \frac{\nu}{1\, {\rm GHz}}\right)^{-2} \; 
\frac{\rm DM}{\rm 10^5\; pc\; cm^{-3}}\; {\rm s}\, ,
\label{DeltatDM}
\end{equation}
where $\nu_p=(n_ee^2/\pi m_e)^{1/2}=8.98\cdot 10^3n_e^{1/2}$~Hz (cgs units). As is discussed in~\cite{FRB}, 
plasma effects with ${\rm DM} = 776.2(5)$~cm$^{-3}$~pc 
could be responsible for the entire $\Delta t_{12}^{\rm 
FRB}$ that was measured~\footnote{In~\cite{FRB}
a different method has been used to obtain the DM value. 
However, for this letter it is enough 
to compare the arrival times of these two frequencies, which reproduces quite accurately 
the result of~\cite{FRB}.}.
 There are contributions to the DM of this extragalactic object from
{the free electron density}  
in the host galaxy, estimated to be $\sim 37$~cm$^{-3}$~pc, from the Milky Way and its halo, estimated to be
$219$~cm$^{-3}$~pc, and the intergalactic medium (IGM). Subtracting the other contributions, the IGM
contribution to the DM was estimated to be $\simeq 520$~cm$^{-3}$~pc, with uncertainties $\sim 38$~cm$^{-3}$~pc
from the modelling of the Milky Way using NE2001~\cite{NE2001}~\footnote{For limitations of NE2001, see~\cite{NE2001critic}.} and $\sim 100$~cm$^{-3}$~pc from inhomogeneities in the IGM.
The ${\rm DM_{IGM}}$ contribution to the dispersion delay~(\ref{DeltatDM}) for a source at red shift $z$
can be expressed in terms of the density fraction ${\rm\Omega_{IGM}}$ of 
ionized baryons~\cite{IGM}:
\begin{equation}
{\rm DM_{IGM}}\; =\; \frac{3cH_0\Omega_{\rm IGM}}{8\pi Gm_p}H_e(z)\, ,
\label{D_M}
\end{equation}
where $H_0$ is the present Hubble expansion rate, $G$ is the Newton constant, $m_p$ is the proton mass, and the factor
\begin{equation}
H_e (z) \; \equiv \; \int_0^z \frac{(1 + z^\prime) d z^\prime}{\sqrt{\Omega_\Lambda + (1 + z^\prime)^3 \Omega_m}} \, ,
\label{He}
\end{equation}
takes proper account of the time stretching in~(\ref{DeltatDM}) and evolution of the free-electron density 
due to the cosmological expansion~\cite{IGM,IGM2,stretch1,stretch2}.
The relation~(\ref{D_M})  was used in~\cite{FRB} to estimate the density of {ionized} baryons in the IGM:
$\Omega_{\rm IGM}^{\rm FRB} = 0.049 \pm 0.013$, assuming that the helium fraction in the IGM has the cosmological 
value of 24\%. We also assume that the present cosmological constant density fraction $\Omega_\Lambda = 0.714$
and the present matter density fraction $\Omega_m = 0.286$, and set 
the reduced Hubble expansion rate,
$h_0 \equiv H_0/(100$~km~s$^{-1}$~Mpc$^{-1}) = 0.69$~\cite{WMAP}.
This measurement of $\Omega_{\rm IGM}$ is 
quite 
compatible with the density
expected within standard $\Lambda$CDM cosmology~\cite{WMAP}: 
$\Omega_{\rm 
IGM}^{\rm\Lambda CDM} = 0.041 \pm 0.002$.

The measurement of $\Delta t_{12}^{\rm FRB}$  can also be used to constrain the photon mass.
For this purpose, we note that the difference in distance covered by two particles emitted by an object at a red shift $z$ 
with velocity difference $\Delta u$ is
\begin{equation}
\Delta L = H_0^{-1}\int_0^z \frac{\Delta u d z^\prime}{\sqrt{\Omega_\Lambda + (1 + z^\prime)^3 \Omega_m}}.
\label{distance1}
\end{equation}
In case of the cosmological propagation of two massive photons with energies 
$E_2>E_1$ the velocity difference is
\begin{equation}
\Delta u_{m_{\gamma}} = \frac{m_{\gamma}^2}{2(1+z)^2}\left(\frac{1}{E_1^2}-\frac{1}{E_2^2}\right)\, ,
\label{velocity1}
\end{equation}
where the red shifts of the photon energies are taken into account and we use units: $\hbar = c = k =1$. Thus,
difference in arrival times of two photons of different energies from a remote cosmological object
due to a non-zero photon mass can be parametrized as follows:
\begin{equation}
\Delta t_{\rm lag} = \frac{m_{\gamma}^2}{2 H_0} \cdot F(E_1, E_2) \cdot H_{\gamma}(z)+\Delta t_{\rm DM}+b_{\rm sf}(1+z)\, ,
\label{arrivalT1}
\end{equation}
where $F(E_1, E_2) \equiv \left(\frac{1}{E_1^2}-\frac{1}{E_2^2}\right)$,
\begin{equation}
H_{\gamma} (z) \; \equiv \; \int_0^z \frac{d z^\prime}{(1 + z^\prime)^2\sqrt{\Omega_\Lambda + (1 + z^\prime)^3 \Omega_m}} \, ,
\label{Hg}
\end{equation}
and we include in (\ref{arrivalT1}) the contribution $\Delta t_{\rm DM}$ to 
the time delay due to plasma effects and a possible, 
generally unknown, source time lag $b_{\rm sf}$ in the source frame. 
Inverting (\ref{arrivalT1}) and transforming to
experimental units $F_{\rm GHz}(\frac{\nu_1}{1\rm GHz},\frac{\nu_2}{1\rm GHz})$  
and expressing all time measurements in seconds
we arrive at
\begin{equation}
m_{\gamma}=(1.05\cdot 10^{-14}\, {\rm eV s^{-1/2}})\sqrt{\frac{h_0}{F_{\rm GHz}H_{\gamma}}
(\Delta t_{\rm lag}-\Delta t_{\rm DM}-b_{\rm sf}(1+z))}\, .
\label{photonMass1}
\end{equation}
The most conservative bound 
\begin{equation}
m_\gamma \; < \; 2.6 \times 10^{-14}~{\rm eV} {\rm c}^{-2} \; (4.6 \times 10^{-50}~{\rm kg}) \, 
\label{conservative}
\end{equation}
 would be obtained if the entire DM of FRB 150418 were due to
$m_\gamma \ne 0$, i.e., $\Delta t_{\rm lag}\lesssim\Delta t_{12}^{\rm FRB}$, 
$\Delta t_{\rm DM}=0$ and $b_{\rm sf}=0$ in (\ref{photonMass1}).  
However, this approach is probably too {conservative}, and a very reasonable
assumption would be to subtract from the ${\rm DM}_{\rm IGM}^{\rm FRB}$  the IGM contribution 
corresponding 
to $\Omega_{\rm IGM}^{\rm\Lambda CDM}$. In this case, since 
the 95\%~CL estimate of the IGM dispersion measure
is ${\rm DM}_{\rm IGM(2\sigma )}^{\rm FRB}\simeq 520\pm (2\cdot 138$)~cm$^{-3}$~pc~\cite{FRB},   
one should assume,
according to (\ref{D_M}) and (\ref{DeltatDM}), that $\Delta t_{\rm lag}\lesssim 0.82$~s at the 95\% CL,
 $\Delta t_{\rm DM}\approx 0.45$~s and $b_{\rm sf}=0$ in (\ref{photonMass1}).  
In this case, one would find 
\begin{equation}
m_\gamma \; < \; 1.8 \times 10^{-14}~{\rm eV} {\rm c}^{-2} \; (3.2 \times 10^{-50}~{\rm kg}) \, 
\label{reasonable}
\end{equation} 
at the 95\% CL.~\footnote{Similar
bounds were given in~\cite{Meszaros}, which we received while working on this paper.} These bounds are
much stronger than those obtained from GRBs~\cite{GRBmg} and AGNs, and are getting within shouting
distance of the {PDG limit \cite{ry97,PDG,ry07}}.
We regard this as the most reasonable interpretation of the data on FRB 150418.

The question then arises, how much the FRB limit could be improved in the future?

The DM of FRB 150418 has been measured with an accuracy of 0.1\%, but the uncertainties
in subtracting the contributions from the host galaxy, the IGM and the Milky Way amount to $> 20$\%.
In particular, uncertainties associated with inhomogeneities in the IGM approach 20\%, dwarfing
uncertainties associated with $\Omega_{\rm IGM}$, which approach 5\%, and 
in modelling the
Milky Way~\cite{NE2001,NE2001critic}, which exceed 5\%. We doubt that the corresponding uncertainties for other FRBs
could soon be reduced to the 0.1\% level of the FRB 150418 DM measurement,
and consider that a plausible objective may be to constrain the sum of 
DM$_{\rm IGM}$
and a possible {photon-mass} effect for any given FRB with an accuracy of 
10\%.~\footnote{In this
respect we are considerably less optimistic than the authors of~\cite{Meszaros}.} One way to
improve the sensitivity to $m_\gamma$ may be to use data from FRBs at different redshifts. As we
discuss below, the relative contributions of the IGM and a photon mass vary with the redshift $z$,
and the sensitivity to $m_\gamma$ is greater for FRBs with smaller redshifts. A
hypothetical 10\% measurements of the non-host and non-Milky Way
contributions to the DM of a FRB with $z = 0.1$ would yield a prospective sensitivity
to $m_\gamma = 6.0 \times 10^{-15}$~eV c$^{-2}$ ($1.1 \times 10^{-50}$~kg).

As already commented, the frequency dependences of the IGM and $m_\gamma$ 
effects, Eqs.
(\ref{DeltatDM}) and  (\ref{photonMass1}), are similar, but the degeneracy 
between them is broken by the different $z$ dependences
of $H_e$ (\ref{He}) and $H_\gamma$ (\ref{Hg}).
In particular, we note the $m_\gamma$ effect gains in relative more importance at smaller $z$ because of the
difference between the powers of $(1 + z^\prime)$ in the integrands of $H_e$ and $H_\gamma$. In practice,
if in the future a statisticaly relevant sample of FRBs at different redshifts is observed one might use
the parametrization (\ref{arrivalT1}) to recover the intrinsic time lag of every source $i$ from the sample as 
\begin{equation}
b_{\rm sf}^i  = \frac{1}{(1+z^i)}(a_{\gamma}^i\cdot F(E_1,E_2)\cdot H_{\gamma}(z)+
\Delta t_{\rm DM}^i-\Delta t_{\rm lag}^i)\, .
\label{bsf}
\end{equation}
Assuming identical origins for the FRBs, one could optimize the set of $b_{\rm sf}^i$ with respect to $a_{\gamma}^i$
and ${\rm \Omega_{\rm IGM}^i}$ ($\Delta t_{\rm DM}^i$), separating the non-zero photon mass contribution out from
the plasma effect. The optimization can be performed on a basis of some estimator: a simple one could be just
a minimization of the RMS of $b_{\rm sf}^i$.~\footnote{A variant of such algorithm has been used in~\cite{SN} 
for neutrino mass estimations from a supernova signal.} 

\begin{figure}
\centering
\vspace{-1cm}
\includegraphics[scale=0.55]{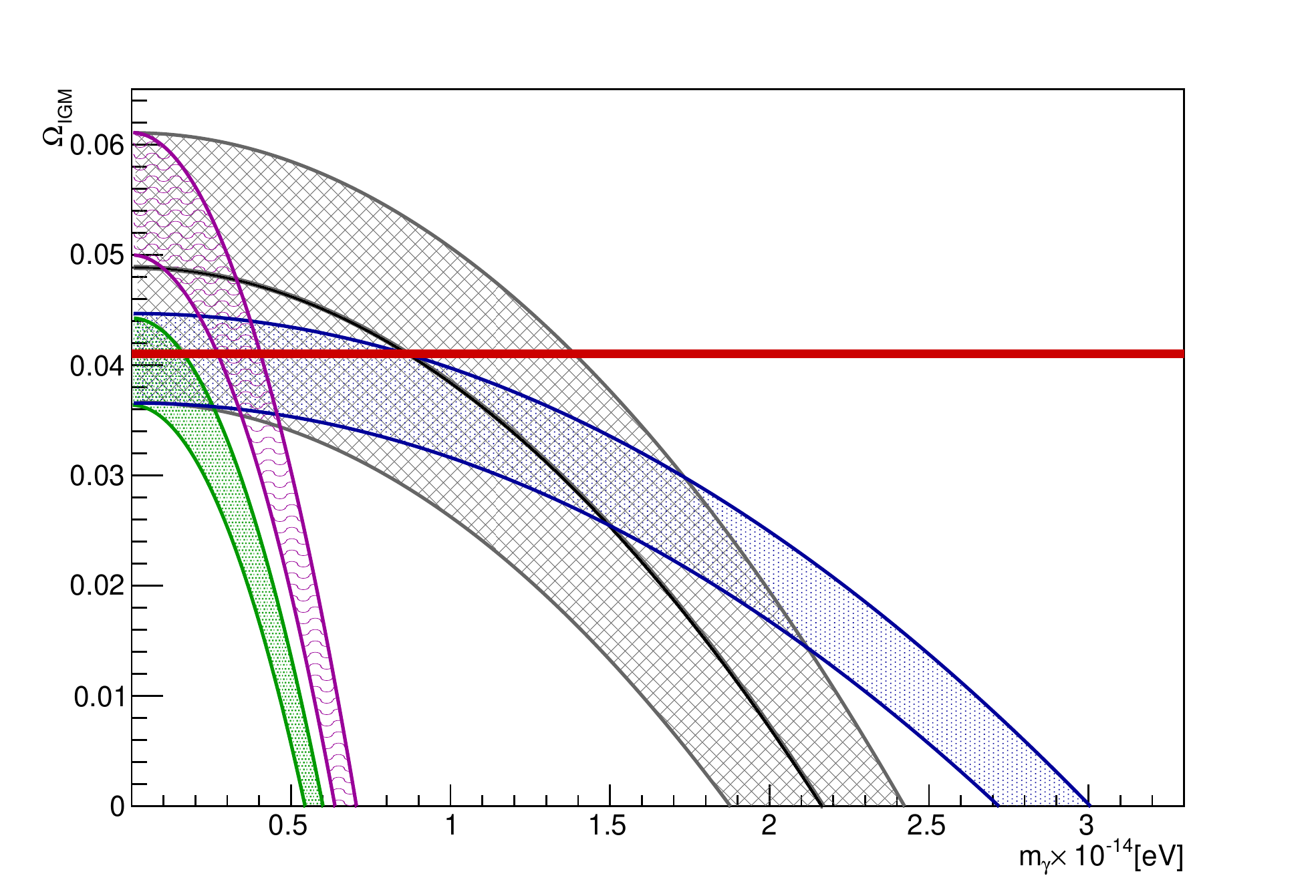}
\vspace{-0.4cm}
\caption{\it The $(m_\gamma, \Omega_{\rm IGM})$ plane, showing
as a thin horizontal red band the $\Lambda$CDM expectation that 
$\Omega_{\rm IGM} = 0.041 \pm 0.002$, 
a curved grey shaded band representing the FRB 150418 constraint as discussed in the text,
and other bands representing the impacts of hypothetical future 10\% measurements
of the extragalactic DM for FRBs with redshifts $z = 0.1$ (green and mauve) and $z = 1.0$ (blue).}
\label{fig:mIGM}
\end{figure}

As discussed above, we consider that future measurements of the non-host galaxy and
non-Milky Way contributions to the DMs of other FRBs at the 10\% level may be feasible
objectives. Accordingly, we have made a first assessment of their possible future impacts on the
photon mass limit. Figure~\ref{fig:mIGM} displays an $(m_\gamma, 
\Omega_{\rm IGM})$ plane,
featuring as a thin horizontal band the $\Lambda$CDM expectation that
${\rm \Omega_{\rm IGM}^{\Lambda CDM}}$. The other curves have the forms
\begin{equation}
m_{\gamma}=A\sqrt{B-C}\,
\label{curves}
\end{equation}
that follows from (\ref{photonMass1}), where $A$ is a numerical pre-factor determined by 
the factor $H_{\gamma}(z)$ of an object, the term $B$ represents an observed time lag in terms of 
intergalactic DM 
\begin{equation}
B=(103.1\, {\rm s})\cdot\frac{\rm DM_{\rm IGM}^{obs}}{10^5\, {\rm pc ~
cm^{-3}}}
\label{B}
\end{equation}  
and $C$ defines the fraction of an actual contribution of the ionized plasma effect
to the observed time lag relative to the prediction of the standard $\rm\Lambda CDM$ model
for a given object 
\begin{equation}
C=\Delta t_{\rm IGM}\cdot\frac{\Omega_{\rm IGM}}{\Omega_{\rm IGM}^{\rm\Lambda CDM}}.
\label{C}
\end{equation}
The curves in Figure~\ref{fig:mIGM} assume an ionization fraction 0.9 but allow $\Omega_{\rm IGM}$ to be a free parameter.  
The curved grey shaded band shows the FRB 150418 constraint discussed above, at the 68\% CL, which implies 
$A=2.96\cdot 10^{-14}\, {\rm eV\cdot s^{-1/2}}$, 
${\rm DM_{\rm IGM}^{obs}}={\rm DM_{\rm IGM}^{FRB}}$
and $\Delta t_{\rm IGM}=0.45\, {\rm s}$. 
The intersection of
this band with the $\Omega_{\rm IGM} = 0$ axis corresponds to the (overly?) conservative 95\% CL limit 
(\ref{conservative}) and its intersection with the $\Lambda$CDM band for $\Omega_{\rm IGM}$
corresponds to the `reasonable' 95\% CL bound (\ref{reasonable}).

The Figure also displays other bands, showing the potential impacts of hypothetical 10\% measurements
of the extragalactic DM for FRBs with redshift $z = 0.1$ (green and mauve) 
and $z = 1.0$ (blue).~\footnote{The
low luminosities of FRBs would render them difficult to detect at larger $z$.} The hypothetical
$z = 0.1$ green band has the same central value as expected for 
${\rm\Omega_{\rm IGM}^{\Lambda CDM}}$
and a massless photon, for which case $A=1.97\cdot 10^{-14}\, {\rm eV\cdot s^{-1/2}}$,
${\rm DM_{\rm IGM}^{obs}}={\rm 83\, pc\cdot cm^{-3}}$ 
and $\Delta t_{\rm IGM}=0.086\, {\rm s}$ have been used in 
(\ref{B}) and (\ref{C}). ~\footnote{For all 
hypothetical sources a 10\% unceartanty
in ${\rm DM_{\rm IGM}^{obs}}$ is applied.} 
The $z = 1.0$ blue band has been calculated with $A=4.60\cdot 10^{-14}\, {\rm eV\cdot s^{-1/2}}$,
 ${\rm DM_{\rm IGM}^{obs}}={\rm 903\, pc\cdot cm^{-3}}$ and $\Delta t_{\rm 
IGM}=0.94\, {\rm s}$ applied in (\ref{B}) and (\ref{C}) . 
The hypothetical $z = 0.1$ mauve band has the same upper limit on $\Omega_{\rm IGM}$
as the FRB 150418 measurement and differs from the green one in having ${\rm 
DM_{\rm IGM}^{obs}}={\rm 103\, pc\cdot cm^{-3}}$
used in (\ref{B}) and \ref{C}).
As expected, we see that a 10\% measurement of an FRB with $z = 0.1$ yielding the 
expected central value (green band) would impose a more stringent
constraint on $m_\gamma$, namely
\begin{equation}
m_\gamma \; < \; 6.0 \times 10^{-15}~{\rm eV} {\rm c}^{-2} \; (1.1 \times 10^{-50}~{\rm kg}) \, .
\label{future}
\end{equation}
if one (very conservatively) allows any $\Omega_{\rm IGM} \ge 0$, strengthening to $< 3 \times 10^{-15}$~eV c$^{-2}$
for ${\rm\Omega_{\rm IGM}^{\Lambda CDM}}$. Alternatively, we see that 
consistency of the green 
band with the FRB 150418 constraint would
require $m_\gamma < 2.5 \times 10^{-15}$~eV c$^{-2}$, without any assumption on $\Omega_{\rm IGM}$. 

We also see that consistency between a `high' measurement from an FRB with $z= 0.1$ (mauve band) and
an `expected' measurement from an FRB with $z = 1.0$ (blue band) would be consistent with 
${\rm\Omega_{\rm IGM}^{\Lambda CDM}}$ only
if one requires a non-zero $m_\gamma \in [2.5, 4.0] \times 10^{-15}$~eV c$^{-2}$.
These are just examples of possible future developments in the interpretation of possible DM measurements from
future FRBs with measured redshifts, and specifically how the effects of the IGM and a photon mass could in principle
be distinguished. Significant improvements on these estimated sensitivities would require more careful estimates
of possible reductions in the uncertainties in DM$_{\rm IGM}$, in 
particular, and would benefit
from a combined analysis of a larger number of FRBs.

For completeness, we mention another way to bound $m_\gamma$ using radio emissions,
namely by comparing the the arrival time of radio afterglow and $\gamma$-ray emission from 
a GRB. The most promising example seems to be GRB~071109 which was observed~\cite{Frail} to
exhibit a radio afterglow at 8.46~GHz about 0.03~d after its $\gamma$-ray 
emission.~\footnote{Other
GRBs have less sensitivity, because there were larger delays before their afterglows were detected.}
Although the redshift of this GRB was not measured, assuming that its redshift lies within the range $z \in [0.1, 5]$, 
we find an upper limit on the photon mass 
$m_\gamma \lesssim 2.8 \times 10^{-11}$~eV ${\rm c}^{-2}$ ( = 5.0 $\times 
10^{-47}$~kg).~\footnote{Here 
we assume simultaneous emission of the radio waves and $\gamma$ rays, which may not be the case. 
If the radio waves were emitted before the $\gamma$ rays (foreglow), any delay due to the
photon mass would be masked by the earlier time of emission.} The weakness
of the limit compared to the FRB limit discussed earlier is due to the much larger time delay before
the observation of the radio afterglow. Whilst this
limit is not competitive with the FRB limit given above or the limit currently quoted by the PDG, this 
GRB afterglow method has the interest of
involving a different type of astrophysical modelling. Moreover, it has potential for future
improvement, {\it e.g.}, if one could use lower-frequency waves and/or observe an afterglow sooner after the
parent GRB, and particularly if time structure in the radio emissions analogous to those in the
$\gamma$-ray emissions could be detected.

We finish our discussion with come comments and speculations. 
The present lack of redshift measurements for other FRBs is an obstacle for obtaining a more robust 
upper bound on the photon mass. However, one could also reverse the logic used above for FRB 150418 and,
assuming the expected cosmological density of the IGM and the upper limit on the photon mass derived from FRB 150418,
estimate the redshifts of other observed FRBs. Their redshift distribution might help pin down their origins.
Another option would be to use gravitational lensing, which would become frequency dependent
in the presence of a photon mass~\cite{OtherReviews}. The lensing is independent of the distance from the source, and 
a photon of mass $m_\gamma$ and energy $E$ from a source of mass $M$ would be
gravitationally deflected by an angle $\theta = \frac{4M\, {\rm G}}{R\, {\rm c}^2}\, \Big( 1 + \frac{m_\gamma^2\, c^4}{2 E_\gamma^2}\Big)$, 
for a photon of energy $E$ (or frequency $\nu = E/h$), where $R$ is the size of the celestial body and $G$
is the gravitational constant. In \cite{OtherReviews}, the photon-mass deflection $\Delta \theta$
was set equal to the difference between the value observed for some celestial object, e.g., the Sun,
and the standard theoretical case for massless photon, thereby obtaining an upper bound 
$m_\gamma \lesssim h \nu {\rm c}^{-2}\, \sqrt{2\Delta \theta/\theta_0} $, where $\theta_0 =  \frac{4M\, {\rm G}}{R\, {\rm c}^2}$ 
is the standard massless photon deflection. Limits of the order of $m_\gamma \lesssim 10^{-44}$\, kg can be obtained this way. 
Conversely, using upper bounds of the photon mass obtained from other methods like the FRBs discussed here 
would remove one uncertainty in the predictions for expected deflection angles,
sharpening the use of comparisons with observations to constrain better the properties of lensing objects. 

\section*{Acknowledgements}

The research of JE and NM was supported partly by the London Centre for Terauniverse Studies (LCTS), 
using funding from the European Research Council via the Advanced Investigator Grant 26732, 
and partly by the STFC Grant ST/L000326/1. JE thanks Dale Frail for useful communications
and the Universidad de Antioquia,
Medell\'in, for its hospitality while this work was initiated, using Grant FP44842-035-2015 from Colciencias (Colombia). The work of AS
was supported partly by the US National Science Foundation under Grants No.PHY-1205376
and No.PHY-1402964.

\end{document}